\documentclass[conference]{IEEEtran}
\IEEEoverridecommandlockouts
\usepackage{cite}
\usepackage{amsmath,amssymb,amsfonts}
\usepackage{multirow}
\usepackage{algorithmicx,algorithm}
\usepackage{subfigure}
\RequirePackage{booktabs}
\usepackage{fancyhdr}
\usepackage{fancyhdr} 

\usepackage{graphicx}
\usepackage{textcomp}
\usepackage{xcolor}
\def\BibTeX{{\rm B\kern-.05em{\sc i\kern-.025em b}\kern-.08em
    T\kern-.1667em\lower.7ex\hbox{E}\kern-.125emX}}

\begin{document}
\pagestyle{empty}  

\title{A Rate Control Algorithm for Video-based Point Cloud Compression\\
	\thanks{Corresponding author: Wei Gao (gaowei262@pku.edu.cn). This work was supported by Ministry of Science and Technology of China - Science and Technology Innovations 2030 (2019AAA0103501), Natural Science Foundation of China (61801303 and 62031013), Guangdong Basic and Applied Basic Research Foundation (2019A1515012031), Shenzhen Fundamental Research Program (GXWD20201231165807007-20200806163656003), Shenzhen Science and Technology Plan Basic Research Project (JCYJ20190808161805519), Open Projects Program of National Laboratory of Pattern Recognition (NLPR) (202000045), and CCF-Tencent Open Fund (RAGR20200114).}
}

\author{\IEEEauthorblockN{1\textsuperscript{st} Fangyu Shen}
	\IEEEauthorblockA{\textit{School of Electronic and Computer Engineering} \\
		\textit{Peking University Shenzhen Graduate School}\\
		Shenzhen, China \\
		shenfy@pku.edu.cn}
	\and
	\IEEEauthorblockN{2\textsuperscript{nd} Wei Gao}
	\IEEEauthorblockA{\textit{School of Electronic and Computer Engineering} \\
		\textit{Peking University Shenzhen Graduate School}\\
		Shenzhen, China \\
		gaowei262@pku.edu.cn}
}

\maketitle



\pagestyle{fancy}
\fancyhf{} 
\thispagestyle{fancy}
\lhead{\footnotesize \copyright2021 IEEE. Personal use of this material is permitted. Permission from IEEE must be obtained for all other uses, in any current or future media, including reprinting/republishing this material for advertising or promotional purposes, creating new collective works, for resale or redistribution to servers or lists, or reuse of any copyrighted component of this work in other works. DOI: 10.1109/VCIP53242.2021.9675449.}

\begin{abstract}
Video-based point cloud compression (V-PCC) has been an emerging compression technology that projects the 3D point cloud into a 2D plane and uses high efficiency video coding (HEVC) to encode the projected 2D videos (geometry video and color video).  In this work, we propose a rate control algorithm for the all-intra (AI) configuration of V-PCC.  Specifically, based on the quality-dependency existing in the projected videos, we develop an optimization formulation to allocate target bits between the geometry video and the color video. Furthermore, we design a two-pass method for HEVC to adapt to the new characteristics of projected videos, which significantly improves the accuracy of rate control.
Experimental results demonstrate that our algorithm outperforms V-PCC without rate control in R-D performance with just 0.43\% bitrate error.
\end{abstract}


\section{Introduction}
A point cloud is a three-dimensional (3D) data representation that has attracted extensive attention for describing the spatial structure and properties of 3D objects. Each point in point clouds contains rich 3D coordinate information (x, y, z) and optional characteristics such as color~\cite{ref1}.
In recent years, point clouds have been widely employed in various immersive media applications~\cite{ref2}. 
However, thousands of points represent a 3D object, which is a problem for hardware storage and network transmission, prompting the development of the high efficiency point cloud compression (PCC).

The Moving Picture Expert Group (MPEG) develops two distinct compression technologies: geometry-based PCC (G-PCC) and video-based PCC (V-PCC)~\cite{ref3}. Different from G-PCC, which compresses the point cloud in its original domain, V-PCC projects the position and color attributes of the 3D point cloud into different 2D patches~\cite{addref1}. The projected 2D patches are then placed in different 2D images with the size of W × H, which make up 2D videos (geometry video, color video)~\cite{ref4}. In this case, we can make full use of mature 2D video compression frameworks such as HEVC~\cite{addref2} to compress the projected 2D videos.

Due to the limited bandwidth in actual transmission, rate control (RC) is an indispensable part of a compression framework~\cite{ref5,ref6,ref7}. Nonetheless, the RC algorithm is under-explored for V-PCC. To the best of our knowledge, there are few studies on V-PCC rate control. In~\cite{ref8}, the authors calculate the QPs of the geometry and color videos by solving a bit allocation problem. To provide more precise rate control, Liu \textit{et al.}~\cite{ref9}  take advantage of the rate-distortion (R-D) characteristics of seven regions to calculate the QP for each region. The adjustment of QP in the work above is still rough. To this end,~\cite{ref10} employs the rate control mechanism in HEVC to refine QP adjustment, in which each frame in the geometry and color videos can be assigned a different QP.

Nevertheless, in~\cite{ref10}, the quality-dependency between the geometry and color videos is not considered in the bit allocation. The following are the primary contributions of this work: (1) We are the first to propose an efficient RC algorithm for AI configuration of V-PCC and validate its good performance on V-PCC test model (TMC2-10.0)~\cite{ref12}. Note that TMC2-10.0 is associated with HEVC reference software (SCM8.8)~\cite{HM}. (2) An optimal bit allocation formulation is proposed by establishing a quality-dependency model and R-D models. (3) A two-pass rate control algorithm for HEVC is designed to improve the accuracy of rate control. Different from other two-pass schemes~\cite{addref3,addref4}, our scheme not only improves the accuracy of the model parameters but also guides the bit allocation of the group-of-pictures (GOP).

The rest of this paper is structured as follows. Section \uppercase\expandafter{\romannumeral2} introduces the proposed novel rate control framework for V-PCC, including an optimal bit allocation algorithm
and a two-pass rate control method for HEVC. The experimental results are presented in Section \uppercase\expandafter{\romannumeral3}. Finally, Section \uppercase\expandafter{\romannumeral4} concludes the paper.

\section{Proposed Rate Control Algorithm for V-PCC}
In V-PCC, the point cloud generates occupancy map (OCC), geometry video (GV), color video (CV), and auxiliary patch information (PATCH) through several techniques such as path and packing. The OCC is encoded with a lossless video encoder, and the PATCH also selects lossless encoding. Accordingly, the consumed bits of the compressed OCC and PATCH are constant. It means that the proposed bit allocation algorithm only works on GV and CV. The two-pass-based RC method is then adopted to adjust GV and CV bits to the allocated target bits. The proposed method is described in full further below.

\subsection{Optimal Bit Allocation Algorithm}

\begin{figure}[t] 
	\centering  
	\subfigtopskip=2pt 
	\subfigbottomskip=2pt
	\subfigcapskip=-5pt 
	\subfigure[$R_{G}-D_{G}$ model]{
		\label{level.sub.1}
		\includegraphics[width=0.40\linewidth]{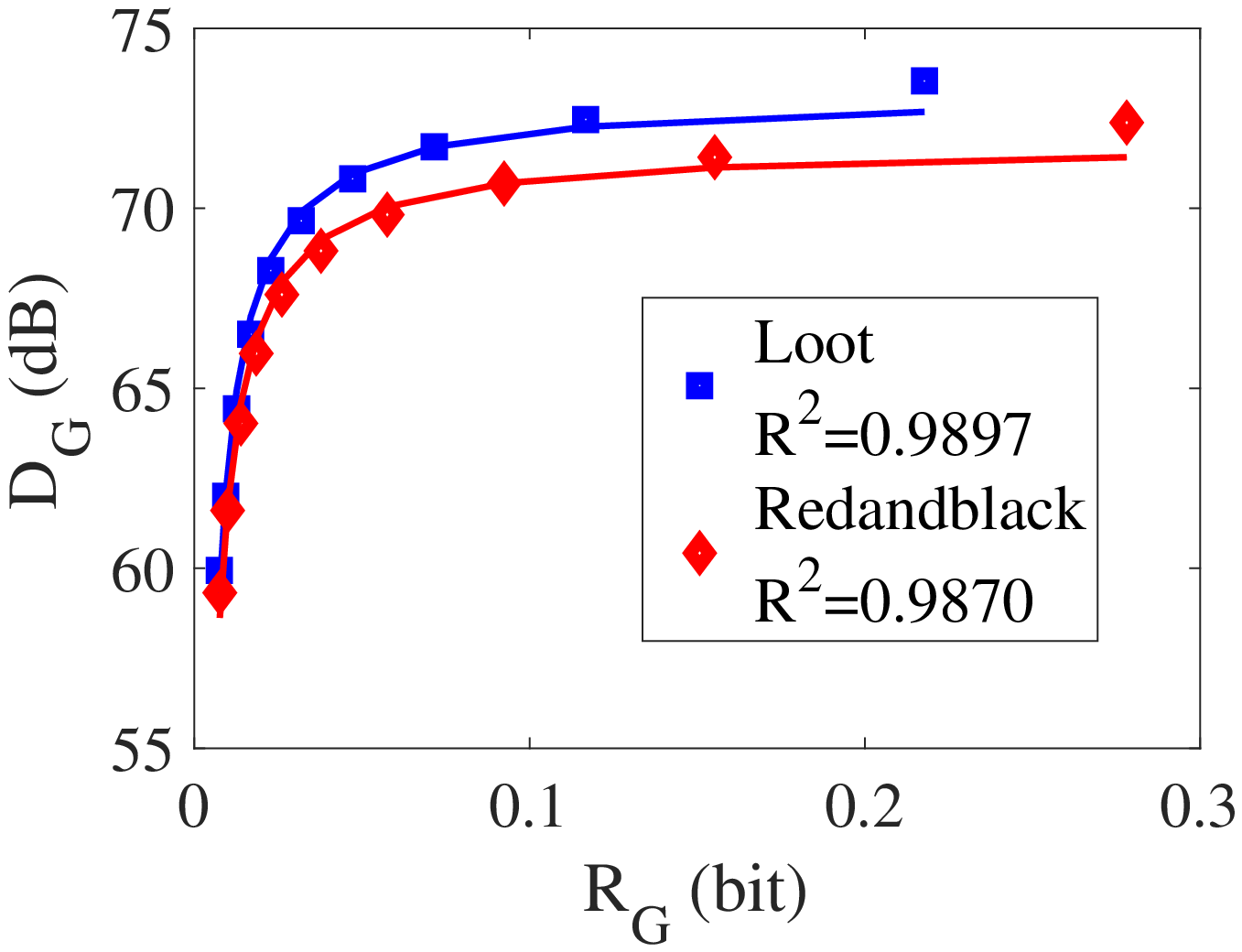}}
	\subfigure[$R_{G}-D_{G}$ model]{
		\label{level.sub.2}
		\includegraphics[width=0.40\linewidth]{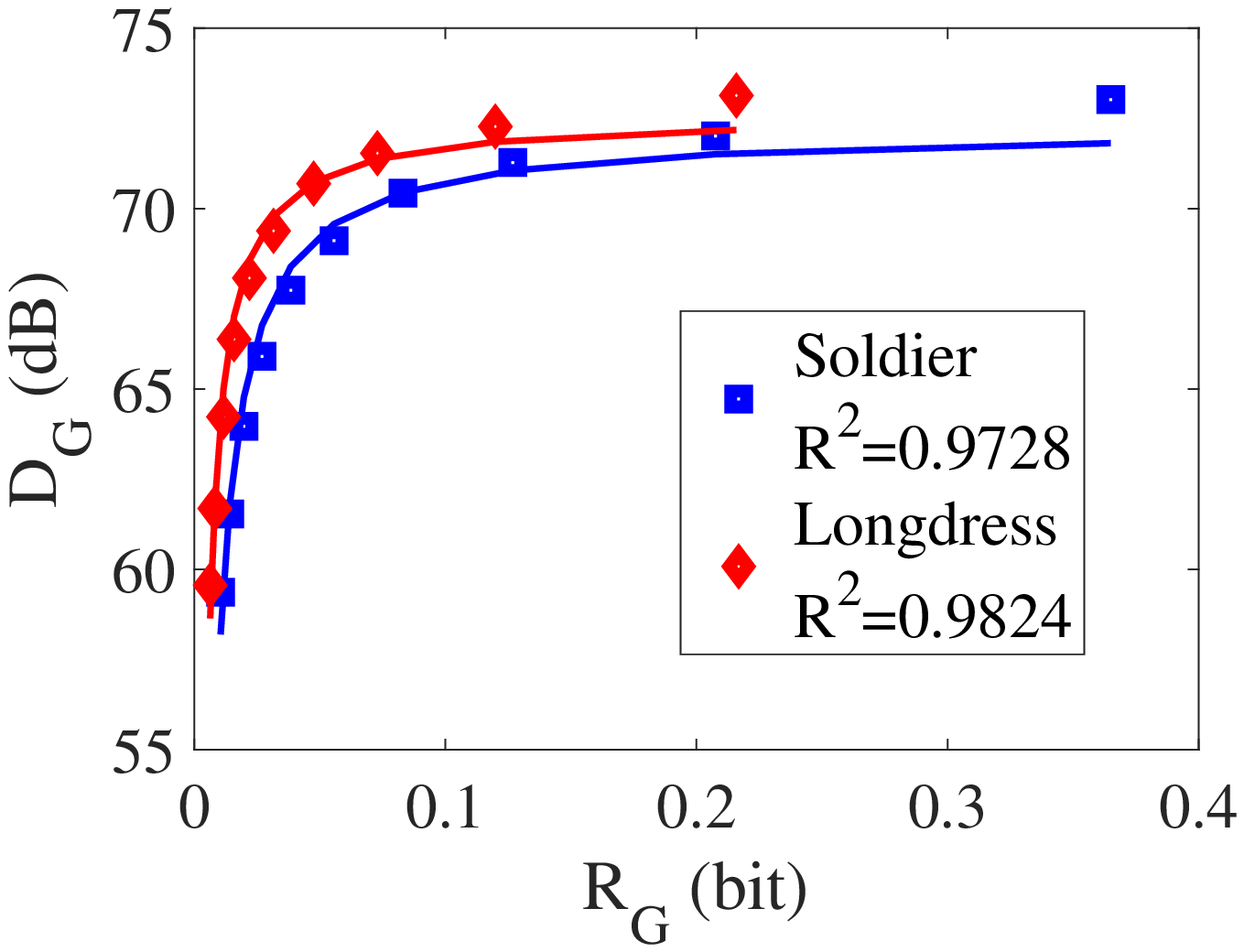}}
	\subfigure[$R_{C}-D_{C}$ model]{
		\label{level.sub.1}
		\includegraphics[width=0.40\linewidth]{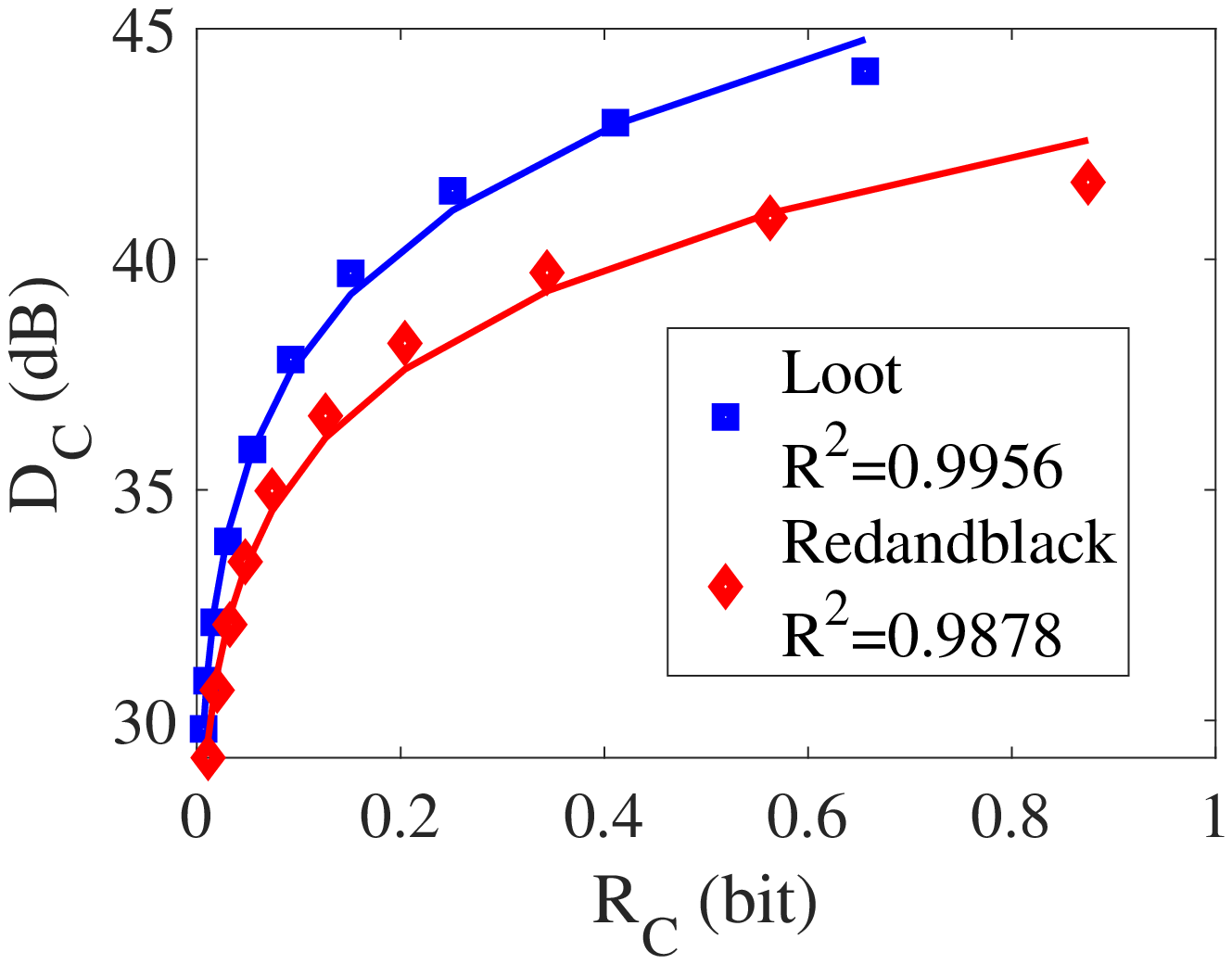}}
	\subfigure[$R_{C}-D_{C}$ model]{
		\label{level.sub.2}
		\includegraphics[width=0.40\linewidth]{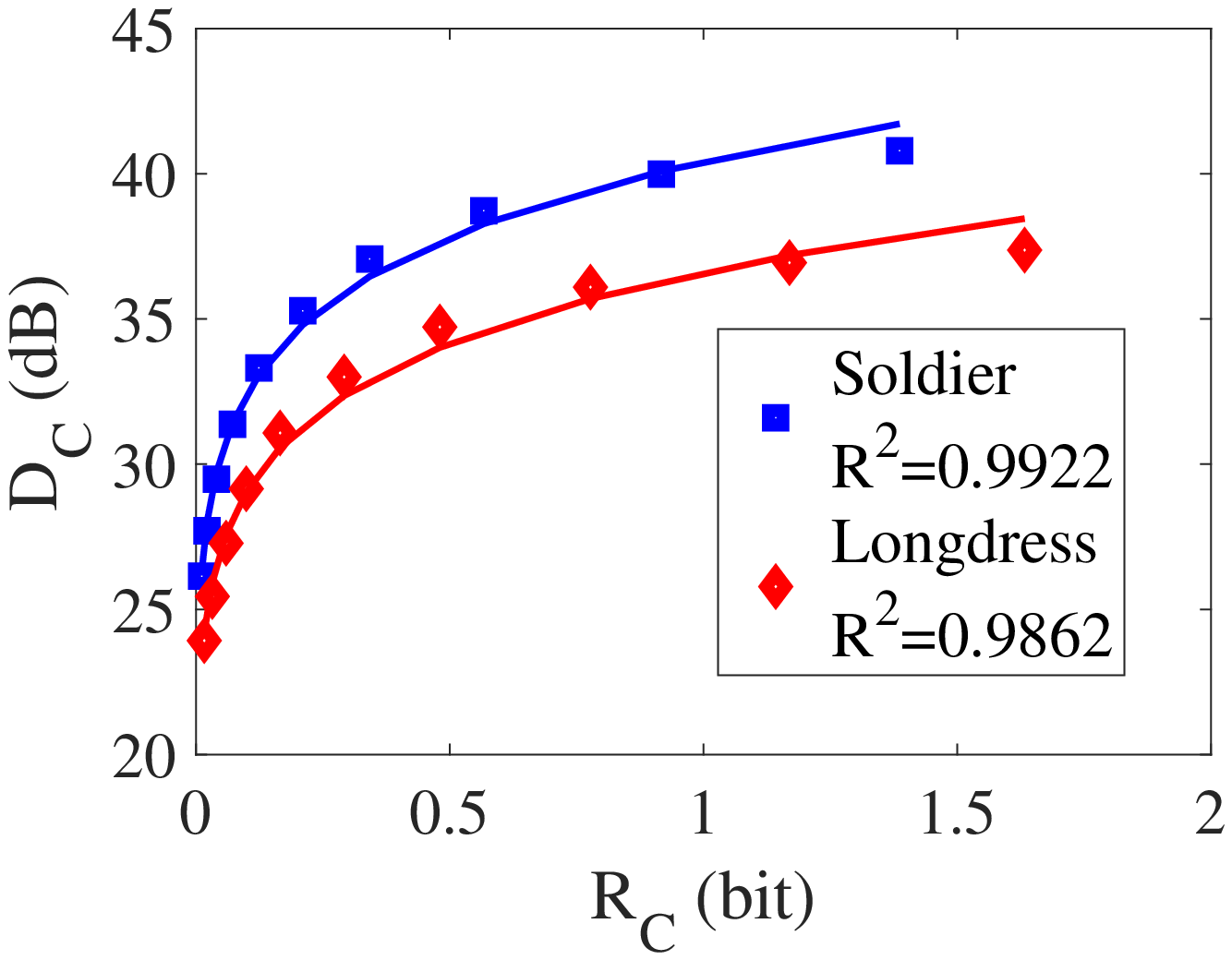}}
	\caption{Illustration of the R-D relationship. Subscripts G and C denote the geometry and color video, respectively.}
	\label{figD-R}
\end{figure}
{\bf Rate-Distortion (R-D) models for  geometry and color videos:} Several symbols are used for simplicity, whose exact meaning can be found in Table \ref{tab1}. Refer to~\cite{ref11}  for the detailed calculation process of distortion metrics $D_{G}$ and $D_{C}$. $D_{G}$ represents D1 in~\cite{ref11} , and $D_{C}$ represents the point-to-point distortion in the Y component. In this paper, $D_{G}$ and $D_{C}$ are converted to PSNR values. The larger the  $D_{G}$  and $D_{C}$ values, the smaller the point cloud coding distortion. Seven point clouds of three classes under different $QP_{G}$ settings ($8, 12,\dots,40, 44$) are encoded to derive the $R_{G}-D_{G}$ model. Because the $QP_{G}$ affects $D_{C}$ and $R_{C}$, when investigating the relationship between $R_{C}$ and $D_{C}$, we set $QP_{C}$ to ($14, 18,\dots,46, 50$) with a fixed $QP_{G}$ ($24$). The results for four point clouds are shown in Fig.~\ref{figD-R}. For the curve fitting, the following proposed R-D models are used,
\begin{eqnarray}\label{dgrg}
D_{G}=a_{g}\cdot R_{G}^{-1}+b_{g},
\end{eqnarray}
\begin{eqnarray}\label{dcrc}
D_{C}=a_{c}\cdot R_{C}^{0.1}+b_{c},
\end{eqnarray}
where $a_{g}$, $b_{g}$, $a_{c}$, $b_{c}$ represent the parameters of the models. Besides, $a_{g}$ is a negative number. On the contrary, $a_{c}$ is a positive number. ${\partial D_{G}}/{\partial R_{G}}$ and ${\partial D_{C}}/{\partial R_{C}}$ are derived as,
\begin{equation}\label{gg}
\frac{\partial D_{G}}{\partial R_{G}}=\theta_{g} \cdot  R_{G}^{-2},
\end{equation}
\begin{equation}\label{cc}
\frac{\partial D_{C}}{\partial R_{C}}=\theta_{c} \cdot  R_{C}^{-0.9},
\end{equation}
where $\theta_{g}$ is equal to $-a_{g}$, and $\theta_{c}$ is equal to $0.1 \cdot a_{c}$. Both $\theta_{g}$ and $\theta_{c}$ are greater than 0. It is worth noting that equations \eqref{gg} and \eqref{cc} will be used in the optimal bit  allocation.
\thispagestyle{empty}
\begin{figure}[t] 
	\centering  
	\subfigtopskip=2pt 
	\subfigbottomskip=2pt
	\subfigcapskip=-5pt 
	\subfigure[$QP_{C}$=22]{
		\label{level.sub.1}
		\includegraphics[width=0.40\linewidth]{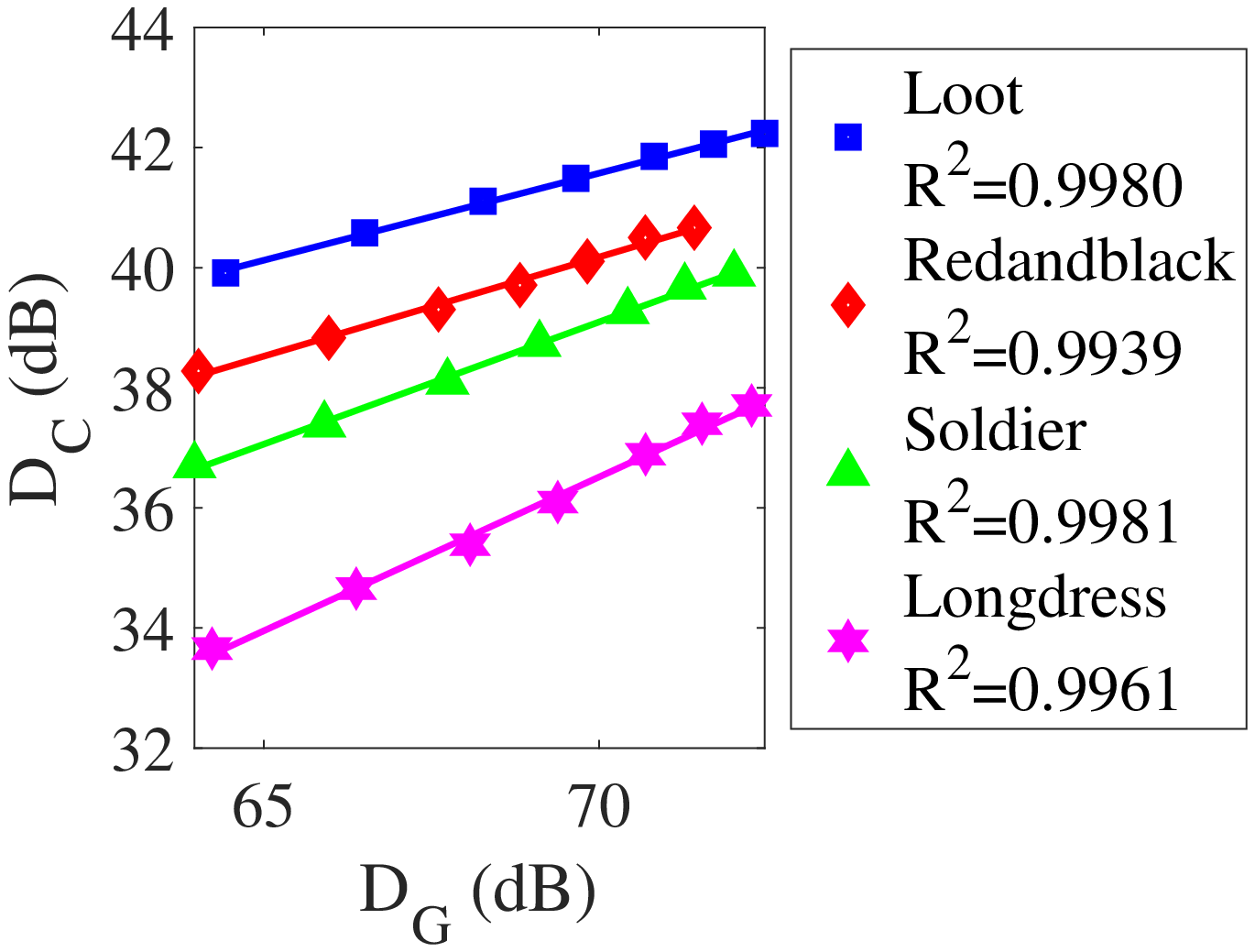}}
	\subfigure[$QP_{C}$=26]{
		\label{level.sub.2}
		\includegraphics[width=0.40\linewidth]{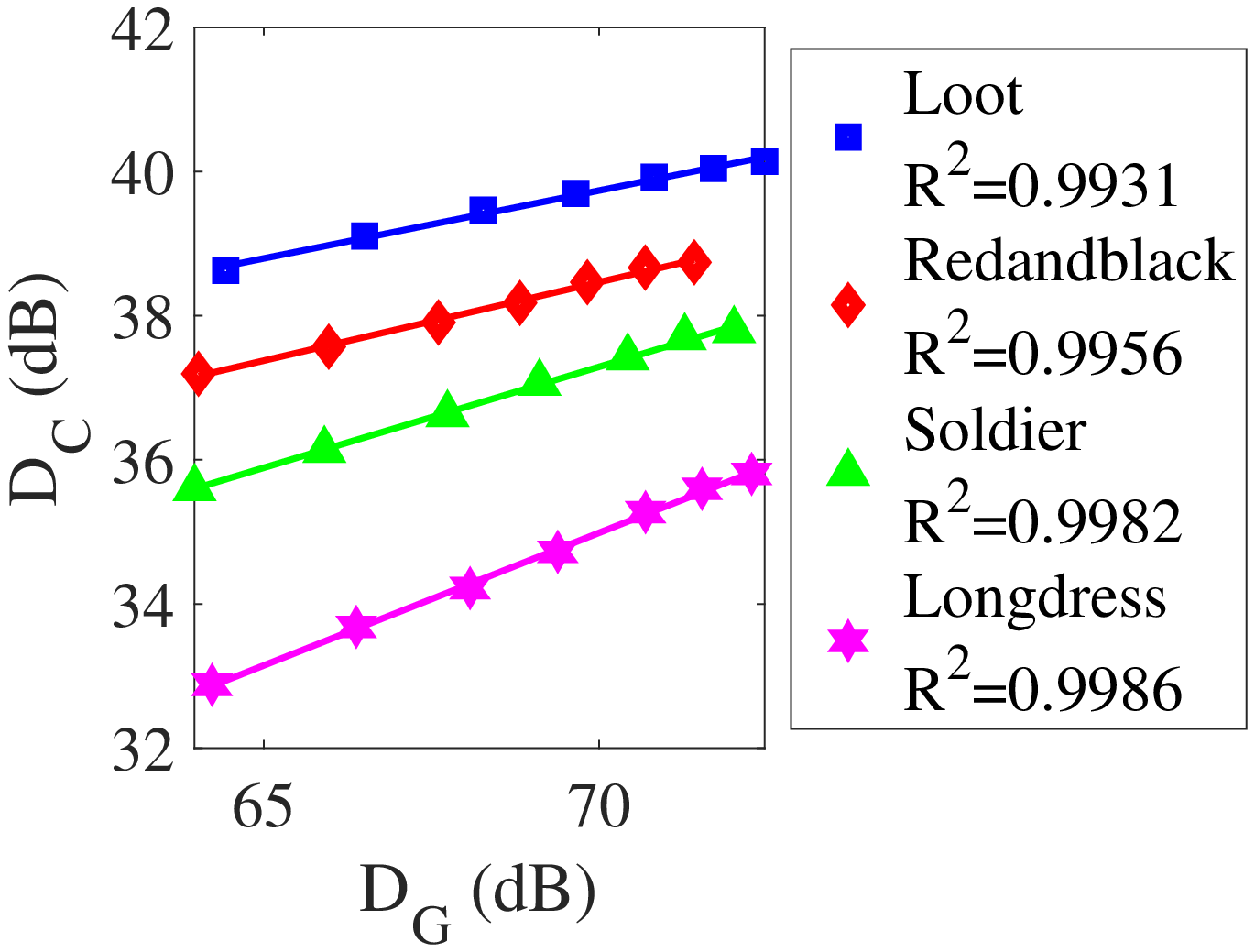}}
	\subfigure[$QP_{C}$=30]{
		\label{level.sub.1}
		\includegraphics[width=0.40\linewidth]{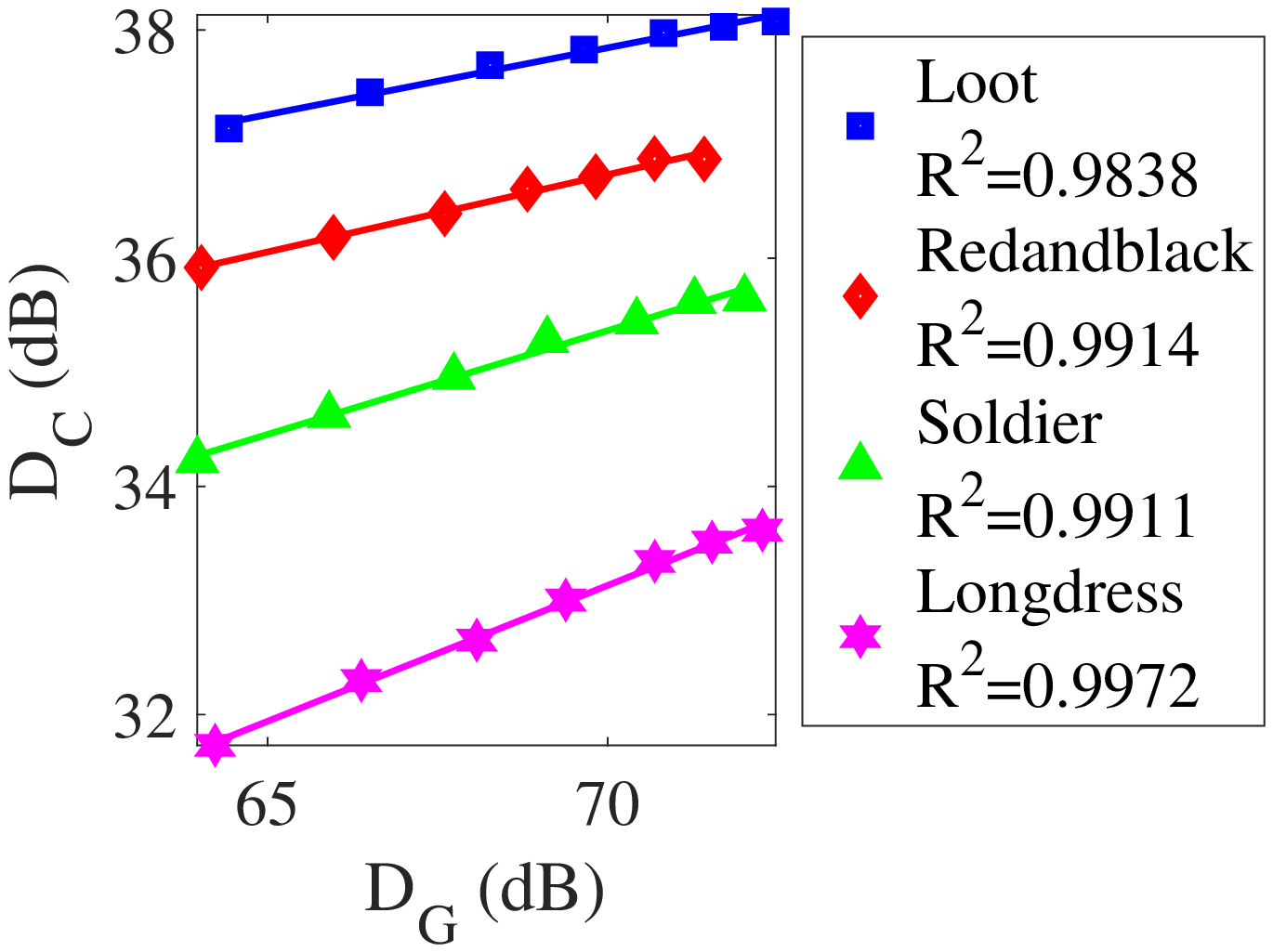}}
	\subfigure[$QP_{C}$=34]{
		\label{level.sub.2}
		\includegraphics[width=0.40\linewidth]{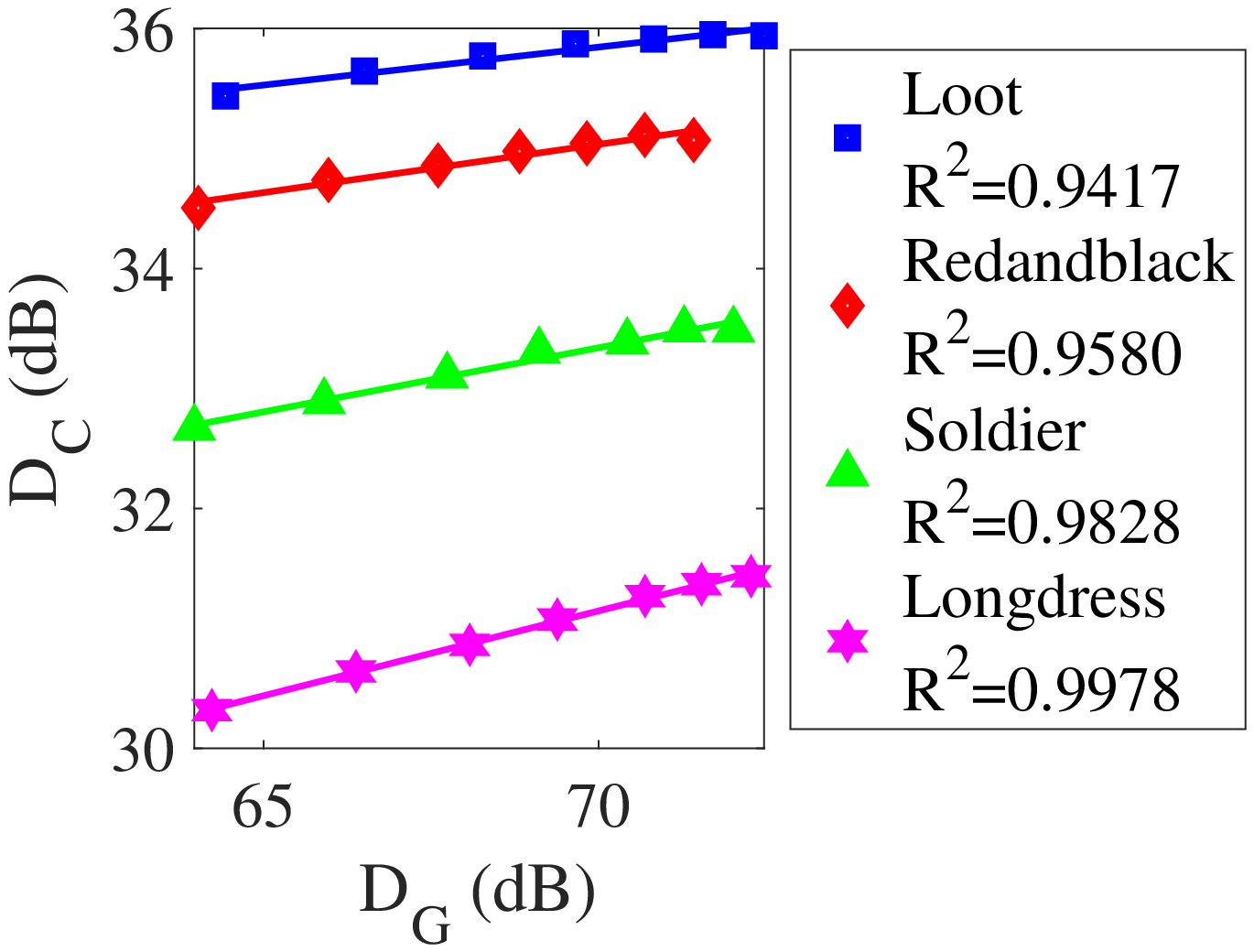}}
	\caption{Illustration of the quality-dependency between $D_{C}$ and $D_{G}$.}
	\label{figyilai}
\end{figure}
The R-squared ($R^{2}$) correlation coefficient is chosen to measure the accuracy of the proposed R-D models. Closer $R^{2}$ to 1 demonstrates higher model accuracy. As shown in Fig.~\ref{figD-R}, most of the $R^{2}$ results are close to 0.98, which implies sufficiently high accuracy of the R-D models.

{\bf Quality-dependency modeling:}
The point cloud is reconstructed from the geometry video, which contains location information. Following reconstructed point clouds, V-PCC recolors each point in point clouds. Thus, the quality of the compressed geometry video has a significant impact on the quality of the color video. We fix the $QP_{C}$ to 22, 26, 30, 34 with various $QP_{G}$ in the range of 12-36 (the change step is 4) to encode seven different point clouds. From Fig.~\ref{figyilai}, an approximated linear relationship between $D_{C}$ and $D_{G}$ for the same $QP_{C}$ can be found,
\begin{eqnarray}\label{Eq1}
D_{C}=\kappa \cdot D_{G}+b,
\end{eqnarray}
where $\kappa$ is a linear coefficient ranging between [0.1, 0.5]. Based on numerous experiments, $\kappa$ is set to 0.3 in this paper.
\begin{equation}\label{yilai}
\frac{\partial D_{C}}{\partial D_{G}}=\kappa.
\end{equation}

Moreover, the $R^{2}$ results of four point clouds are presented in Fig.~\ref{figyilai}, which confirms the accuracy of \eqref{Eq1}.

{\bf Optimal bit allocation algorithm:}
Unlike the 2D compression task, V-PCC has more than one distortion metric, including $D_{G}$ and $D_{C}$. It is necessary to integrate the two metrics to define the overall distortion. In the common test conditions (CTC)~\cite{ref11} of V-PCC, $QP_{G}$ is about five smaller than $QP_{C}$, which means the quality of geometry is far more important than that of color. For this reason, the overall distortion of the point cloud $D_{total}$ can be expressed as:
\begin{equation}\label{Dtotal}
D_{total}=w \cdot D_{G}+D_{C},
\end{equation}
\begin{table}[t]\setlength{\tabcolsep}{6mm}
	\centering
	\caption{Symbols in This Paper}
	\begin{tabular}{ll}
		\toprule
		Symbol	 &  Meaning \\
		\midrule
		$R_{G}$    & Consumed bits of the geometry video \\
		$R_{C}$    & Consumed bits of the color video \\
		$D_{G}$    & Distortion (PSNR) of the geometry video \\
		$D_{C}$    & Distortion (PSNR) of the color video\\
		$QP_{G}$   & QP of the geometry video \\
		$QP_{C}$   & QP of the color video \\
		\bottomrule
	\end{tabular}%
	\label{tab1}%
\end{table}%
where  $w$ is the weight of the geometry video. In this paper, $w$ is set to 25. We set $w$ to 20, 25 and 30, respectively. Experimental results show that when $w$=25\footnote[1]{The acquisition process of the hyperparameter $w$ takes the total point cloud performance as the comparison target. In order to balance the distortions of geometry video and color video, the $w$ value is set as $w'$=0.259 after considering the number of point cloud frames. Due to the limited pages of the VCIP conference, we provide more details here.}, the R-D performance is excellent. Under the limitation of target bits $R_{tar}$, the optimal bit allocation seeks to minimize the total distortion of the point cloud,  which can be derived by:
\begin{equation}\label{min}
\text  { min } \quad  \{-(w \cdot D_{G}+D_{C}) \}\quad s.t. \quad R_{G}+R_{C} \leq R_{tar}.
\end{equation}
Note that $D_{C}$ and $D_{G}$ are PSNR values, which means that the larger the value of $D_{C}$  and $D_{G}$, the smaller the point cloud distortion. Next, we can introduce the Lagrangian multipliers method to solve the constrained optimization problem, 
\begin{equation} \label{mmmm}
\text { min } \{-(w \cdot D_{G}+D_{C})+\lambda\left(R_{G}+R_{C}-R_{tar}\right)\},
\end{equation}
where $\lambda$ is the Lagrangian multiplier for the optimization problem. Taking the derivative of  \eqref{mmmm} for  $R_{G}$ and  $R_{C}$, respectively, we get the following expression:
\begin{equation}\label{qiudqo1}
-w \frac{\partial D_{G}}{\partial R_{G}}-\frac{\partial D_{C}}{\partial R_{G}}+\lambda=0,
\end{equation}
\begin{equation}\label{qiudqo2}
-w \frac{\partial D_{G}}{\partial R_{C}}-\frac{\partial D_{C}}{\partial R_{C}}+\lambda=0.
\end{equation}
Convert Equations \eqref{qiudqo1} and \eqref{qiudqo2} to \eqref{qiudqo3} and \eqref{qiudqo4}, respectively,
\begin{equation}\label{qiudqo3}
-w \frac{\partial D_{G}}{\partial R_{G}}-\frac{\partial D_{C}}{\partial D_{G}} \cdot \frac{\partial D_{G}}{\partial R_{G}}+\lambda=0,
\end{equation}
\begin{equation}\label{qiudqo4}
-w \frac{\partial D_{G}}{\partial D_{C}} \cdot \frac{\partial D_{C}}{\partial R_{C}}-\frac{\partial D_{C}}{\partial R_{C}}+\lambda=0,
\end{equation}
where $\frac{\partial D_{C}}{\partial D_{G}}$ represents the inﬂuence of the geometry on color, which is discussed in \eqref{yilai}. Combining \eqref{gg}, \eqref{yilai} and \eqref{qiudqo3}, we can get:
\begin{equation}\label{rg}
R_{G}=\sqrt{\frac{(w+\kappa) \cdot \theta_{g}}{\lambda}}.
\end{equation}
Similarly, $\frac{\partial D_{G}}{\partial D_{C}}$ represents the inﬂuence of the color on geometry which is equal to 0. Combining \eqref{cc} and \eqref{qiudqo4}, we have:
\begin{equation}\label{rc}
R_{C}=\left(\frac{\theta_{c}}{\lambda}\right)^{\frac{10}{9}}.
\end{equation}
Combining \eqref{min}, \eqref{rg} and \eqref{rc}, the optimization problem will be formulated as:  
\begin{equation}\label{final}
\sqrt{\frac{(w+\kappa) \cdot \theta_{g}}{\lambda}} + \left(\frac{\theta_{c}}{\lambda}\right)^{\frac{10}{9}} \leq R_{tar}.
\end{equation}
\thispagestyle{empty}
\begin{table}[t]\setlength{\tabcolsep}{3mm}
	\centering
	\caption{Bitrate Error of Original RC Method in HEVC}
	\begin{tabular}{lccccc}
		\toprule
		\textbf{Point cloud } & Loot   & Soldier & Queen & Longdress \\
		\textbf{Bitrate error} & 36.84\%  & 93.22\% & 31.29\% & 24.01\% \\
		\bottomrule
	\end{tabular}%
	\label{tab2}%
\end{table}%

\begin{algorithm}[t]
	\caption{Iterative Solution Search for $\lambda$ } 
	\hspace*{0.02in} {\bf Input:} \\
	\hspace*{0.18in} The  range: [${\lambda}_{min}$, ${\lambda}_{max}$]\\
	\hspace*{0.18in} The initialized value: ${\lambda}_{init}$\\
	\hspace*{0.18in} The target bits for geometry and color videos: $R_{tar}$ \\
	\hspace*{0.02in} {\bf Output:} 
	The best Lagrange multiplier ${\lambda}_{comp}$
	\begin{algorithmic}[1]
		\State${\lambda}_{comp}$=${\lambda}_{init}$
		\State{\bf for }{Iter in [1, Itermax] }  {\bf do }
		\State{\quad \quad Calculate the consumed bits R(${\lambda}_{comp}$)}
	    \State{\quad \quad  \bf if }{R(${\lambda}_{comp}$) $>$ $R_{tar}$}  {\bf then }
		\State \quad \quad \quad \quad ${\lambda}_{min}$=${\lambda}_{comp}$
		\State\quad \quad \quad \quad  ${\lambda}_{comp}$=(${\lambda}_{comp}$+${\lambda}_{max}$) $/$ 2
		\State{\quad \quad \bf else }
		\State\quad \quad \quad \quad ${\lambda}_{max}$=${\lambda}_{comp}$
		\State\quad \quad \quad \quad ${\lambda}_{comp}$=(${\lambda}_{comp}$+${\lambda}_{min}$) $/$ 2
		\State{\bf return } ${\lambda}_{comp}$
	\end{algorithmic}
\end{algorithm}
The analytical solution of \eqref{final} is still difficult to obtain. Note that  $\theta_{g}$ and $\theta_{c}$ are fixed by the average value of seven point clouds and greater than zero, which indicates that \eqref{rg} and \eqref{rc} are monotonic functions of $\lambda$. According to the monotonical relationship, an iterative solution search method is introduced to deal with the optimization formulation. Algorithm 1 summarizes the core concepts of search strategy. The range of  $\lambda$ is  [$0.01$, $100$], and the max iteration number is set to 20. The initialization of $\lambda$ is $4$.
After the Lagrange multiplier $\lambda$ is obtained, the $R_{G}$ and $R_{C}$ can be computed by \eqref{rg} and \eqref{rc}. At this point, the whole optimal bit allocation algorithm flow is completed.

\subsection{Two-Pass Rate Control for Video Encoder}
According to  the proposed optimal bit allocation algorithm, we test the original RC algorithm~\cite{ref5} of HEVC reference software (SCM8.8)~\cite{HM} in V-PCC. The experimental results in Table \ref{tab2} suggest the poor performance of the original RC algorithm. Two main reasons for the poor performance are found by analyzing experimental data.

1) Due to the lack of previous knowledge, the initialization parameters of the R-${\lambda}$ model cannot be very accurate.

\begin{figure}[t]
	\centering 
	\vspace{-0.20cm}
	\subfigtopskip=2pt 
	\subfigbottomskip=2pt
	\subfigcapskip=-5pt 
	\centerline{\includegraphics[width=8cm]{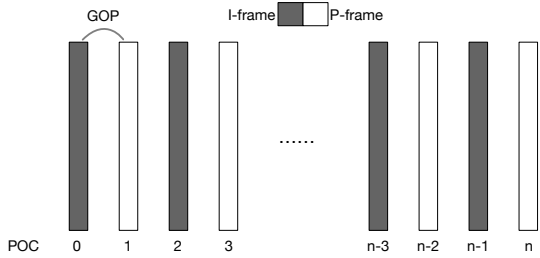}}
	\caption{IPIP..IPIP coding structure for geometry video and color video.}
	\label{figIP}
\end{figure}

2) In the generation process of 2D videos, each patch is projected into two images: a near image and a far image, which solves the problem of multiple points projected to the same point~\cite{ref1}. In view of this characteristic, HEVC adopts the IPIP ... IPIP (IP) coding structure  shown in  Fig.~\ref{figIP} to encode the projected videos. Since the IP coding structure is quite different from the traditional ones, including  All-Intra (AI), Random Access (RA), and Low Delay (LD), the existing RC algorithm in SCM8.8 can not sufficiently  adapt to this new structure. 

To address these shortcomings, a two-pass rate control algorithm is designed, which not only improves the accuracy of the initialization parameters of the model but also guides the bit allocation of IP structure well.

{\bf Make the model initialization parameters more accurate:} In our two-pass RC method, we employ the R-Q model instead of the R-${\lambda}$ model, which can be represented as:
\begin{eqnarray}\label{RQm}
R=a\cdot QP^{b},
\end{eqnarray}
where $a$ and $b$ denote parameters of the model. We convert equation \eqref{RQm} to the logarithm domain,
\begin{eqnarray}\label{RQlg}
\ln R=\ln a + b\cdot \ln QP.
\end{eqnarray}

Before the actual encoding, four different QPs are used to pre-encode the first frame of the geometry and color videos. Later, the consumed bits $R$ are recorded, and four different pairs of ($R_{i}$, $QP_{i}$; $i=1, 2, 3, 4$) can be used to approximate the linear relationship parameters ($\ln a$, $b$) by applying the least squares method~\cite{5348301}. Eventually, the accurate initialization of R-Q model parameters ($a$, $b$) can be achieved. 

\begin{table}[t]
	\centering
	\caption{Proportion of I-frame Consumed Bits in A GOP with Different QPs for Geometry Video}
	\begin{tabular}{lccccc}
		\toprule
		\multicolumn{1}{c}{\multirow{2}[4]{*}{\shortstack{Point\\Cloud}}} & \multicolumn{5}{c}{QPs} \\
		\cmidrule{2-6}          & 44    & 36    & 28    & 20    & 12 \\
		\midrule
		Loot  & 97.45\% & 95.70\% & 90.49\% & 82.23\% & 72.29\% \\
		Redandblack & 96.97\% & 92.26\% & 89.45\% & 81.89\% & 69.39\% \\
		Soldier & 98.27\% & 95.22\% & 90.29\% & 82.81\% & 71.49\% \\
		Queen & 98.52\% & 91.65\% & 87.39\% & 79.79\% & 68.03\% \\
		Longdress & 97.30\% & 93.08\% & 88.93\% & 82.43\% & 71.40\% \\
		Basketball & 98.81\% & 93.00\% & 89.56\% & 82.68\% & 73.65\% \\
		Dancer & 97.45\% & 93.52\% & 88.87\% & 83.37\% & 73.81\% \\
		\midrule
		\textbf{Average}   & \textbf{97.82}\% & \textbf{93.49}\% & \textbf{89.28}\% & \textbf{82.17}\% & \textbf{71.44}\% \\
		\bottomrule
	\end{tabular}%
	\label{Iframe}%
\end{table}%
{\bf Guide bit allocation:} As shown in  Fig.~\ref{figIP}, an I-frame and a P-frame form an IP GOP.  As shown in Table \ref{Iframe}, we collect the ratio of bits consumed by I-frame in a GOP under different encoding QPs of various sequences. The proportion of I-frame consumed bits is more than 70\% and this proportion is related to the encoding QP. Therefore, $R_{i}$ ($i=1, 2, 3, 4$) obtained in pre-encoding can guide the bit allocation in each IP GOP. By comparing the target bits of each GOP with $R_{i}$ ($i=1, 2, 3, 4$), the closest $R_{i}$ can be found. The larger the QP corresponding to the closest $R_{i}$, the more bits are allocated for the I-frame in the IP GOP. For instance, if the QP corresponding to the closest $R_{i}$ is 28, 89.28\% of the GOP bits are allocated to the I-frame.
\thispagestyle{empty}
\section{Experimental Results}
To confirm the good performance of the proposed RC method, we implement it on TMC2-10.0~\cite{ref12} and the corresponding SCM8.8~\cite{HM}. The anchor is the V-PCC without rate control, called the Fixed-QP method. Seven point clouds are encoded following the  CTC~\cite{ref11} of V-PCC and using AI conﬁguration. The number of encoding frames is 32. After that, the obtained bitrates are designated as the target bitrates for our proposed RC algorithm. The bitrate error and R-D performance after comparison are given in Table \ref{tab3}.
\subsection{Bitrate Error}
As shown in  Table \ref{tab3}, the average bitrate error for all point clouds is 0.43\%, which means our RC algorithm achieves accurate rate control. Moreover, the highest bitrate error is only 0.58\%.
\begin{table}[t]\setlength{\tabcolsep}{1mm}
	\centering
	\caption{Performance of The Proposed RC Method over Fixed-QP Method in TMC2-10.0}
	\begin{tabular}{lccccccc}
		\toprule
		\multicolumn{1}{c}{\multirow{3}[4]{*}{\shortstack{Point\\Cloud}}} & \multicolumn{2}{c}{\multirow{2}[2]{*}{\shortstack{Geom.\\BD-TotalRate}}} & \multicolumn{3}{c}{\multirow{2}[2]{*}{\shortstack{Color.\\BD-TotalRate}}}& \multicolumn{1}{c}{\multirow{3}[4]{*}{\shortstack{Overall.\\BD-\\TotalRate}}} & \multicolumn{1}{c}{\multirow{3}[4]{*}{\shortstack{\shortstack{Bitrate\\ Error}}}} \\
		& \multicolumn{2}{c}{} & \multicolumn{3}{c}{}  &       &  \\
		\cmidrule{2-6}          & D1    & D2    & Luma  & Cb    &  Cr   &       &  \\
		\midrule
		Loot  & -5.1\% & -3.1\% & 1.6\% & 1.7\% & 2.8\% & -4.58\% & 0.43\% \\
		\multirow{1}[0]{*}{\shortstack{Redandblack}} & \multirow{1}[0]{*}{-7.1\%} & \multirow{1}[0]{*}{-3.2\%} & \multirow{1}[0]{*}{1.9\%} & \multirow{1}[0]{*}{2.1\%} & \multirow{1}[0]{*}{3.0\%} & \multirow{1}[0]{*}{-6.26\%} & \multirow{1}[0]{*}{0.18\%} \\
		Soldier & 21.3\% & 23.7\% & -0.8\% & -3.8\% & -3.9\% & 19.91\% & 0.39\% \\
		Queen & -11.4\% & -9.2\% & 2.1\% & 3.0\% & 2.7\% & -10.68\% & 0.36\% \\
		Longdress & -48.7\% & -40.6\% & 4.3\% & 8.7\% & 8.6\% & -44.80\% & 0.54\% \\
		Basketball & -2.1\% & -0.6\% & 1.8\% & 1.2\% & 1.2\% & -1.87\% & 0.55\% \\
		Dancer & -4.5\% & -3.0\% & 2.3\% & 2.5\% & 2.5\% & -4.08\% & 0.58\% \\
		\midrule
		\textbf{Average} & \textbf{-8.2\%} & \textbf{-5.2\%} & \textbf{1.9\%} & \textbf{2.2\%} & \textbf{2.4\%} & \textbf{-7.48\%} & \textbf{0.43\%} \\
		\bottomrule
	\end{tabular}%
	\label{tab3}%
\end{table}%
\subsection{R-D Performance}
Bjøntegaard delta (BD)-rate~\cite{ref13} is used to measure the R-D performance of our RC algorithm. The R-D performance is analyzed from the perspectives of geometry, color and geometry+color (overall). Based on \eqref{Dtotal}, the overall R-D performance is computed by $D_{total}$ and total rate.
More bits are allocated to  the geometry video in our bit allocation algorithm because it can produce a more significant R-D performance gain. As shown in Table \ref{tab3}, our algorithm achieves 8.2\% and 5.2\% Geom.BD-TotalRate gains on average in D1 and D2 with only 1.9\% Color.BD-TotalRate loss on average in the Y component. Ultimately, the proposed algorithm achieves 7.48\% Overall.BD-TotalRate gain on average\footnote[2]{The distortion evaluation of V-PCC is divided into two parts: the geometry distortion metric and the color distortion metric. There is no theoretically proper metric for evaluating the overall distortion of the point cloud (especially for the balance of geometry video and color video, which actually requires a complex and comprehensive subjective quality assessment process). As a result, for simplicity, the hyperparameter $w$ was introduced in some existing works to define the overall distortion evaluation metric ($D_{total}$). However, there are currently no fixed weights to evaluate the importance of geometry distortion and color distortion to the overall distortion of point clouds. We should note that the existing settings of $w$ are all empirical, which can trade-off the proportion of geometry distortion and color distortion in the overall point cloud distortion. As we can see, the results in Table \ref{tab3} mean the proposed algorithm achieves a good trade-off between geometry distortion and color distortion. Additionally, we have also provided the Overall.BD-TotalRate performance gain result of 7.48\% as the reference information, and different weights $w$ can result in different Overall.BD-TotalRate. As is widely recognized, the more theoretically proper setting of $w$ demands further investigations. Due to the limited pages of the VCIP conference, we provide more details here.}.

\section{Conclusion}
An efficient rate control algorithm for V-PCC under AI configuration is presented in this paper. In particular, an optimal bit allocation formulation considering quality-dependency is proposed. Meanwhile, an iterative solution search approach is introduced to solve the formulation.
In addition, a two-pass strategy for HEVC is developed to improve the accuracy of rate control. 
Based on the above methods, compared with the Fixed-QP method in V-PCC, the proposed RC algorithm achieves 7.48\% Overall.BD-TotalRate gain on average with only 0.43\% bitrate error.

\thispagestyle{empty}

\end{document}